# High-efficiency diphenylsulfon derivatives-based organic light-emitting diode exhibiting thermally activated delayed fluorescence

Geon Hyeong Lee* and Young Sik Kim[*, **]

*Department of Information Display, Hongik University, Seoul, 121-791, Korea
**Department of Science, Hongik University, Seoul, 121-791, Korea

Novel thermally activated delayed fluorescence (TADF) material with diphenyl sulfone (DPS) as an electron acceptor and 3,6-dimethoxycarbazole (DMOC) and 1,3,6,8-Tetramethyl-9H-carbazole (TMC) as electron donors were investigated theoretically for a blue organic light emitting diode (OLED) emitter. We calculate the energies of the first singlet ($S_1$) and first triplet ($T_1$)-excited states of TADF materials by performing density functional theory (DFT) and time-dependent DFT (TD-DFT) calculations on the ground state using a dependence on charge transfer amounts for the optimal Hartree-Fock percentage in the exchange-correlation of TD-DFT. The calculated $\Delta E_{ST}$ values of TMC-DPS (0.094 eV) was smaller than DMOC-DPS (0.386 eV) because of the large dihedral angles between the donor and accepter moieties. We show that TMC-DPS would have a suitable blue OLED emitter, because it has a large dihedral angle that creates a small spatial overlap between the HOMO and the LUMO and, consequently, the small $\Delta E_{ST}$ and the emission wavelength of 2.82 eV (439.9 nm).

# 1. INTRODUCTION

Since Tang and coworkers reported organic light-emitting devices (OLEDs) with a multi-layer structure, materials and device fabrication have been extensively studied in recent years.[1] Luminescent materials are generally classified into two groups: fluorescent and phosphorescent. In fluorescent OLEDs, only 25% of the excitons can emit light because carrier recombination produces singlet and triplet excitons in a 1:3 ratio through spin statics.[2] Phosphorescent OLEDs based on noble heavy-metal phosphors can produce both singlet and triplet excitons by enhanced spin-orbit coupling (SOC) and can achieve nearly 100% internal quantum efficiency, which corresponds to an external electroluminescent (EL) quantum efficiency of close to 20% without enhanced out-coupling techniques.[3, 4] However, phosphorescent materials containing noble metals, such as Ir(III), Pt(II), and Os(II), are rather expensive and unsustainable. Therefore, a novel method for achieving a high EL efficiency is required.

A newly introduced triplet-harvesting method that uses the thermally activated up-conversion of triplet into singlet states has recently been shown to provide thermally activated delayed fluorescence (TADF) with a high photoluminescence efficiency.[5] Charge transfer (CT) systems confirmed TADF with a small gap between the lowest singlet ($S_1$)- and triplet ($T_1$)-excited states. The $S_1$ level is considerably higher in energy than the $T_1$ level by 0.5–1.0 eV because of the electron exchange energy between these levels. The promotion of TADF requires a small energy gap ($\Delta E_{ST}$) between the $S_1$ and $T_1$ states because the rate of $T_1 \rightarrow S_1$ reverse intersystem crossing (RISC) is inversely proportional to the exponential of $\Delta E_{ST}$.[6] Because $\Delta E_{ST}$ largely depends on the exchange interaction between electrons in molecular orbitals, implying electronic excitation, the spatial overlap between molecular orbitals must be controlled. Thus, the molecular design of TADF emitters involves a careful choice of suitable donor and acceptor units.

In principle, the molecular energy of the lowest singlet ($E_S$) and triplet ($E_t$) excited states can be decided by the orbital energy (E), electron repulsion energy (K), and exchange energy (J) of the two unpaired electrons at the excited states, as shown in Equations (1) and (2). Hence, $\Delta E_{ST}$, which is the difference between $E_S$ and $E_T$, is equal to twice that of J (Equation [3]).

$$E_S = E + K + J \quad (1)$$

$$E_T = E + K - J \quad (2)$$

$$\Delta E_{ST} = E_S - E_T = 2J \quad (3)$$

At $S_1$ or $T_1$, the unpaired two electrons are mainly distributed on the frontier orbitals of the highest occupied molecular orbital (HOMO) and the lowest unoccupied molecular orbital (LUMO), respectively, with the same J value, regardless of the different spin states. Therefore, the exchange energy (J) of these two electrons at HOMO and LUMO can be calculated by Equation (4),

$$J = \int\int \phi_L(1)\phi_H(2)\left(\frac{e^2}{r_1 - r_2}\right)\phi_L(2)\phi_H(1)dr_1 dr_2 \quad (4)$$

where $\Phi_H$ and $\Phi_L$ represent the HOMO and LUMO wave functions, respectively, and e is the electron charge. One important factor in designing molecules with a small $\Delta E_{ST}$ is to have a negligible spatial overlap between the HOMO and the LUMO.[7]

The strategies for obtaining a negligible spatial overlap between the HOMO and the LUMO are: (1) to have a large dihedral angle between the plane of the donor and the connected phenyl rings of the acceptor and (2) to increase the spatial distance between the donor and acceptor constituents with a π-conjugation linker. Previously reported blue TADF emitters showed some strongly localized states with $^3\pi\pi^*$ or $^3n\pi^*$ characters that were lower in energy than the $^3$CT state. This indicated that $\Delta E_{ST}$ can be minimized by adjusting the energy levels of the $S_1$ state and the lowest locally excited triplet state ($^3$LE). Increasing the twist angle between donor and acceptor units limits their electronic interaction, which stabilizes the CT state and increases the energy of the $^3$LE state,[8] although this is insufficient to ensure that the $^3$CT state is lower than $^3$LE. Adachi et al. previously reported a computational prediction for singlet and triplet transition energies of charge transfer compounds using the optimal Hartree-Fock (HF) exchange method to predict the zero-zero energies ($E_{00}$) of the $S_1$, $^3$CT, and $^3$LE levels of molecules.[9]

In this study, we designed novel TADF molecule containing diphenyl sulfone (DPS) as an electron acceptor and 3,6-dimethoxycarbazole (DMOC) and 1,3,6,8-Tetramethyl-9H-carbazole (TMC) as electron donor moieties. We investigated the TADF of the TMC-DPS molecule by the calculated $\Delta E_{ST}$ values using the optimal HF percentage in the exchange-correlation of the time-dependent density functional theory (TD-DFT) and compared to those of DMOC-DPS. In addition, we evaluated the dihedral angles between the planes of the donor and acceptor units to explain the TADF efficiency in terms of their electronic and optical properties.

## 2. COMPUTATIONAL MEHODS

We investigated the factors responsible for the absorption energy and the electron population of molecular orbitals by performing density functional theory (DFT) and time-dependent density functional theory (TD-DFT) calculations on the ground state, using a dependence on the amount of charge transfer (CT) from donor to acceptor for the optimal HF percentage in the TD-DFT exchange-correlation. The geometries in the gas phase were optimized by the DFT method using the B3LYP exchange-correlation function with the 6-31G* basis set in the Gaussian 09 program package. The conformation presented here was the lowest energy conformation, which is the optimal molecular structure of the dyes in the gas phase. The electronic populations of the HOMO and the LUMO were calculated to show the position of electron populations according to the calculated molecular orbital energy diagram. Vertical absorption energies [$E_{VA}$] were calculated by TD-DFT with the B3LYP, MPW1B95, BMK, M062X, and M06HF functionals using 6-31G* basis sets. We determined an optimal HF% (OHF) by using the Multiwfn program to analyze the orbital composition for obtaining a CT amount (q) from the donor to the acceptor; the optimal HF% is proportional to the CT amount with a relation of OHF =42q.

$E_{00}(S_1)$ was obtained using a gap of 0.28 eV between $E_{VA}(S_1,OHF)$ and $E_{00}(S_1)$ from the result of the DMOC-DPS.[10] According to the Franck-Condon principle, the crossing point between absorption and emission corresponds approximately to the $E_{00}$ of the CT transitions, as $E_{VA}(S_1) - E_{00}(S_1) = E_{00}(S_1) - E_{VE}(S_1)$.[11] The value of $E_{00}(S_1)$ was calculated as $(E_{VA}(S_1,OHF) + E_{VE}(S_1,OHF))/2$. The value for $E_{VA}(T_1)$ corresponding to the $^3CT$ or $^3LE$ transitions and calculations for $E_{00}(^3CT)$ and $E_{00}(^3LE)$ were obtained by the proven methods of Adachi et al.[9]

## 3. RESULTS AND DISCUSSION

Carbazole derivative aws designed so the donor unit was changed from DMOC to TMC to enhance the TADF efficiency. TMC was selected as a donor moiety with the two methyl groups in the 1,8-position working to maintain a large dihedral angle through steric repulsion while the two methyl substitutions in the 3,6-position may work to enhance the electrochemical stability of the carbazole ring.[12] We compared these characteristics to those of a reference material DMOC-DPS, which emitted a blue color (445nm).[10] We obtained the energies of the first singlet ($S_1$) and first triplet ($T_1$)-excited states of TADF materials by performing density functional theory (DFT) and time-dependent DFT (TD-DFT) calculations on the ground state using a dependence on charge transfer amounts for the optimal HF percentage in the exchange correlation of TD-DFT. For all the calculations, TD-DFT gave the reasonable emission wavelength of 440 nm of DMOC-DPS in a DCM solution by employing the PCM method. Therefore, we designed TMC-DPS to fabricate a TADF emitter by changing the electron donor from DMOC to TMC. The chemical structures of the materials investigated in the present study are shown in Fig. 1.

The HOMO and the LUMO for this material, calculated with the DFT method using the optimized ground-state molecular geometry, are shown in Fig. 2. The HOMO distribution is largely localized on the cabarzole moiety, while the LUMO is localized on the sulfone moiety, resulting in a small overlap between the HOMO and the LUMO. This TMC-DPS material has a smaller overlap than seen with a previous material of DMOC-DPS. This is because of the change from the DMOC moiety to the TMC moiety. Steric hindrance causes a large dihedral angle between the plane of the TMC electron-donating moieties and the connected DPS electron-accepting moiety. TMC-DPS are receiving greater influence dihedral angle values of 89.7° resulting in a small overlap between the HOMO and the LUMO compared with the dihedral angle of 48.5° for DMOC-DPS. Because $\Delta E_{ST}$ largely depends on the exchange interaction between electrons in molecular orbitals, a smaller spatial overlap between the HOMO and the LUMO leads to a smaller $\Delta E_{ST}$.

Table 1 shows the calculated $\Delta E_{ST}$ and the emission wavelength of the two materials. The calculated $\Delta E_{ST}$ values of TMC-DPS (0.094 eV) was smaller than those of DMOC-DPS (0.386 eV), which is favorable for a RISC process from the $T_1$ to $S_1$ states. The calculated $\Delta E_{ST}$ values indicate that TMC-DPS (0.094 eV) would show higher TADF efficiency. We examined the effect of the steric hindrance between the donor unit and acceptor unit by comparing the $\Delta E_{ST}$ values. As $\Delta E_{ST}$ largely depends on the exchange interaction between electrons in molecular orbitals in Eq. (4), a smaller spatial overlap between the HOMO and the LUMO leads to a smaller $\Delta E_{ST}$. A small $\Delta E_{ST}$ between the $S_1$ and $T_1$ states increases the rate of the $T_1 \rightarrow S_1$ up-conversion, and harvested formally spin-forbidden $T_1$ excitons are shown in Fig. 3. As we expected, the calculated $\Delta E_{ST}$ of TMC-DPS (0.094 eV) was smaller than DMOC-DPS (0.386 eV). In terms of emission wavelength, TMC-DPS showed 2.82 eV (439.9 nm) and DMOC-DPS showed 2.82 eV (439.6 nm), so both materials could reproduce a blue color but only TMC-DPS (0.094 eV) of $\Delta E_{ST}$ was small enough for RISC. Thus TMC-DPS is more suitable for a blue TADF OLED emitter than DMOC-DPS.

## 4. CONCLUSION

In conclusion, we have investigated theoretically novel TADF molecules, DMOC-DPS and TMC-DPS, to enhance TADF efficiencies. An important factor in designing these molecules with a small $\Delta E_{ST}$ is to have a small spatial overlap between the HOMO and the LUMO, with a large dihedral angle between the donor and the acceptor. DMOC-DPS and TMC-DPS show the emission wavelength of 2.82 eV (440 nm). However, the calculated $\Delta E_{ST}$ values of TMC-DPS (0.094 eV) was smaller than those of DMOC-DPS (0.386 eV). Thus, TMC-DPS exhibits a promising blue TADF OLED emitter because of the small $\Delta E_{ST}$ and the large dihedral angle.

**Acknowledgments**

This research was supported by the Basic Science Research Program through the National Research Foundation of Korea (NRF) funded by the Ministry of Education, Science and Technology (2010-0021668).

**Table I.** Calculated data using various exchange-correlation functionals and 6-31G(d) basis set based on B3LYP optimized geometries

| parameter | DMOC-DPS | TMC-DPS |
| --- | --- | --- |
| CT amount (q) | 0.79 | 0.83 |
| Optimal HF% | 33% | 35% |
| dihedral angle (°) | 48.5 | 89.7 |
| $E_{VA}(S_1, OHF)$ (eV) | 3.384 | 3.386 |
| $E_{VE}(S_1)$ (eV, nm) | 2.819 (439.9)<br>2.787 (445.0)* | 2.821 (439.6) |
| $E_{00}(S_1)$ (eV) | 3.101 | 3.103 |
| $E_{00}(^3CT)$ (eV) | 2.862 | 3.136 |
| $E_{00}(^3LE)$ (eV) | 2.715 | 3.009 |
| $\Delta E_{ST}$ (eV) | 0.386 | 0.094 |

*: Experimental data on ref. 10

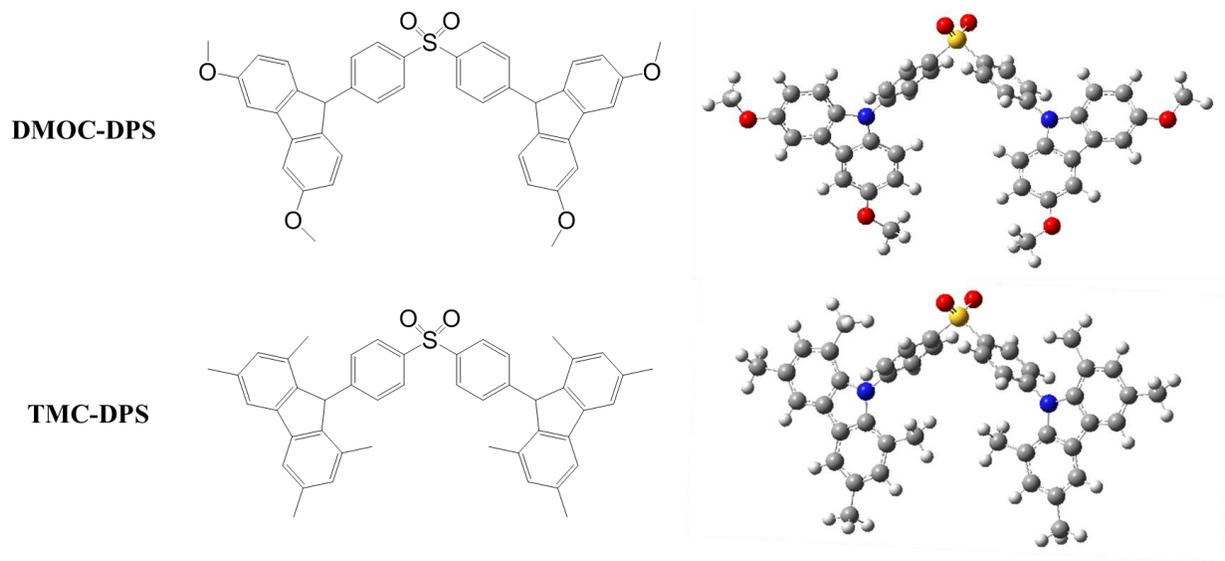

Fig. 1. Chemical structures of DMOC-DPS and TMC-DPS.

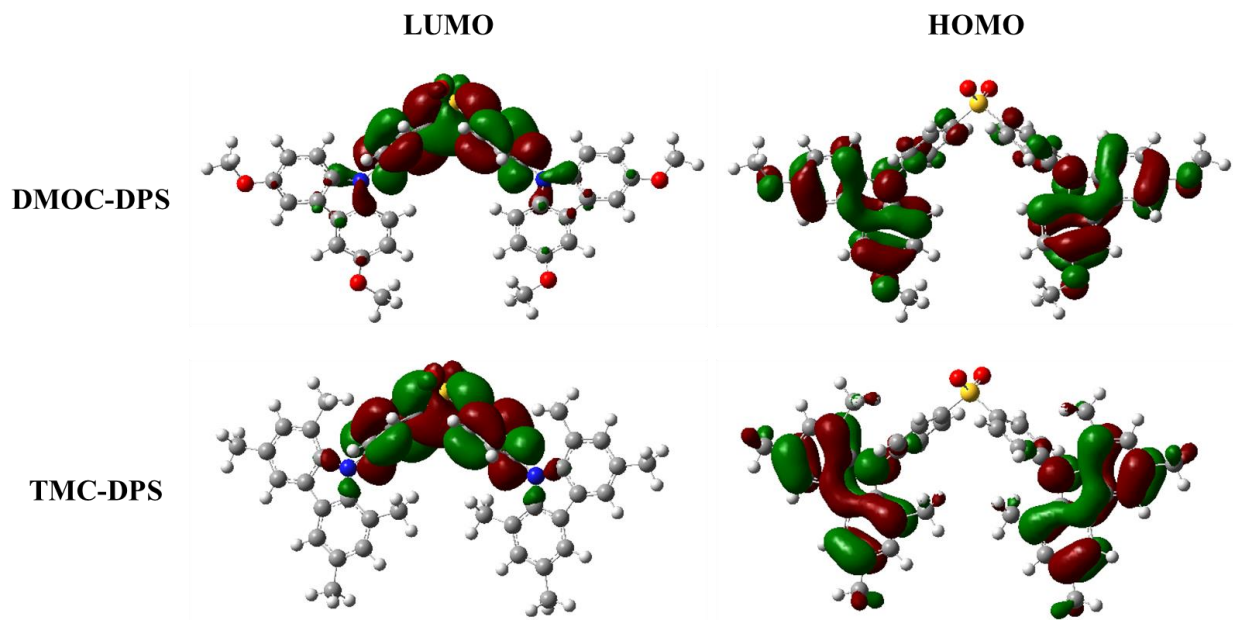

Fig. 2. Schematic drawing of HOMO and LUMO orbital energies for DMOC-DPS and TMC-DPS.

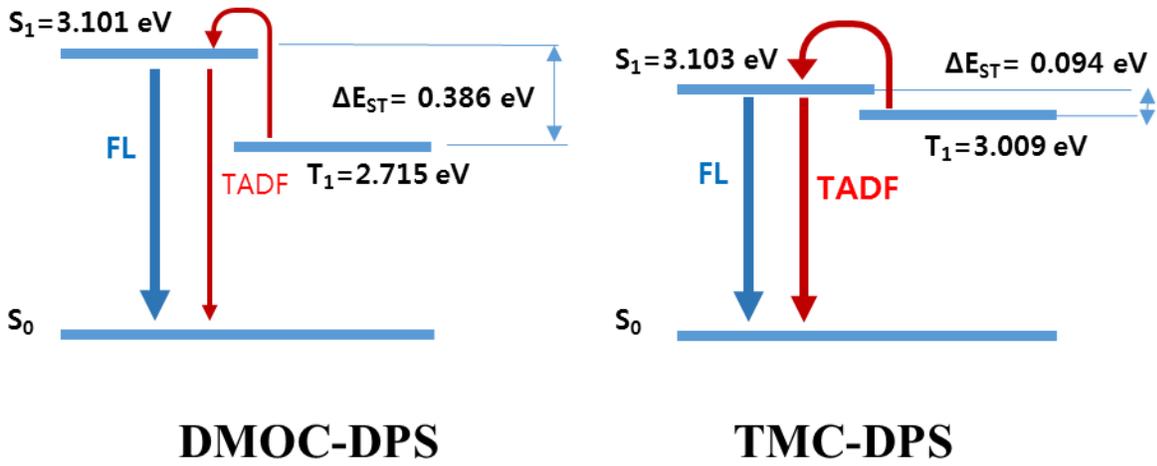

Fig. 3. Schematic energy diagram for DMOC-DPS and TMC-DPS.